\documentstyle[aps,prb,eqsecnum,ifthen,epsfig,multicol]{revtex}
\begin{document}
\draft \title{Zero-bias anomaly in two-dimensional electron layers and
  multiwall nanotubes}

\author{E. G. Mishchenko$^{1,2,3}$ and A. V. Andreev$^{1,2}$ }
\address{{}$^1$ Bell Labs, Lucent Technologies, 600 Mountain Ave, Murray
  Hill, NJ 07974} \address{{}$^2$Department of Physics, University of
  Colorado, CB 390, Colorado 80309-0390} \address{{}$^3$ L. D. Landau
  Institute for Theoretical Physics, Russian Academy of Sciences, Kosygin
  2, Moscow 117334, Russia}

\maketitle

\begin{abstract}{The zero-bias anomaly in the dependence of the
    tunneling density of states $\nu (\epsilon)$ on the energy $\epsilon$
    of the tunneling particle for two- and one-dimensional multilayered
    structures is studied.  We show that for a ballistic two-dimensional
    (2D) system the first order interaction correction to DOS due to the
    plasmon excitations studied by Khveshchenko and Reizer is partly
    compensated by the contribution of electron-hole pairs which is twice
    as small and has the opposite sign.
    For multilayered systems the total correction to the density of states
    near the Fermi energy has the form $\delta \nu/\nu_0 = \mbox{max}
    (\vert \epsilon \vert, \epsilon^*)/4\epsilon_F$, where $\epsilon^*$ is
    the plasmon energy gap of the multilayered 2D system.
    In the case of one-dimensional conductors we study multiwall nanotubes
    with the elastic mean free path exceeding the radius of the nanotube.
    The dependence of the tunneling density of states energy,
    temperature and on the number of shells is found. }

\end{abstract}
\begin{multicols}{2}

\section{Introduction}

Electron-electron interaction results in a singular suppression of the
tunneling (single-particle) density of states at the Fermi surface of
low-dimensional metallic systems.~\cite{AA} The effect, known as the
zero-bias anomaly, was first discussed by Altshuler and
Aronov~\cite{AA79} 
for diffusive systems with a short range interaction, and by Altshuler,
Aronov and Lee~\cite{AAL} for the Coulomb interaction. In the
two-dimensional (2D) case the correction to the density of states (DOS) is
double-logarithmic,\cite{AAL} 
$\delta \nu/\nu_0 \sim \ln{(\vert
  \epsilon \vert \tau)}\ln{(\Delta/\epsilon)}$, 
where $\tau$ is the impurity scattering time,
$\epsilon$ is the energy of a tunneling electron, counting from
the Fermi energy, and $\Delta$ is a high-frequency
cut-off related to the Coulomb screening,
$1/\tau \ll  \Delta \ll \epsilon_F$, $\epsilon_F$ being the Fermi 
energy.  Zuzin~\cite{Z} showed that the second
logarithm in this formula is cut off at low energies for the
experimental setup of a 2D electron plane screened by a metal
shield.  Rudin, Aleiner and Glazman~\cite{RAG} generalized the
theory of zero-bias anomaly to incorporate the ballistic energies
$\epsilon > 1/\tau$, and argued that the correction actually has 
the form, $\delta \nu/\nu_0 \sim -\ln^2{(\epsilon_F/\vert
\epsilon\vert )}$.
Khveshchenko and Reizer~\cite{KhR} analyzed the contribution of
collective electron excitations, 2D plasmons, to the tunneling DOS
and obtained an additional correction $\delta \nu/\nu_0 = (\vert
\epsilon\vert -\epsilon_F)/2\epsilon_F$, less singular near the
Fermi surface but dominant in the wide range of energies.

In the present paper we consider the
interaction correction to the tunneling DOS of a different system,
consisting of a number of identical periodically spaced
two-dimensional electron layers. We also neglect the possibility
of electron tunneling between the layers. While analyzing such a
setup we primarily have in mind high-T$_c$ materials, which are
attracting considerable interest with respect to the properties of
electron-electron interactions. The electron transport in these
materials is extremely anisotropic, and the inter-layer tunneling
amplitude in some crystals 
can be as weak as\cite{ILVYu} $0.05 - 2 ~ K$.
Even though the motion of electrons tunneling from the tip
of the tunneling electron microscope (TEM) is confined to the
outermost layer only, the presence of internal layers is important
as they participate in the screening of the Coulomb interaction.

We also consider a similar though different
tunneling geometry which is realized in multi-wall carbon nanotubes
(MWNT). A typical MWNT consists of a few graphite monolayer sheets rolled
concentrically into cylinders with radius $R \sim 10~ nm$. At zero doping
they can be either metals or semiconductors, depending on the helical
arrangement of the carbon hexagons.  In the measurements of the tunneling
DOS the tunneling current propagates through the outermost shell\cite{B1,B2}
while the inter-shell tunneling is suppressed. Depending on the degree of
disorder the transport around the elastic mean free path $l$ can be shorter
or longer than the radius of the nanotube corresponding to the diffusive
($l<R$) or ballistic ($R>l$) motion along the circumference of the
nanotube.  For energies $\epsilon$ exceeding the inverse time of
propagation around the circumference the zero-bias anomaly is described by
2D formulas. At lower energies a crossover to the regime of a quasi-one
dimensional (quasi-1D) conductor is realized.

The zero-bias anomaly in diffusive quasi-1D conductors was first
considered by Altshuler and Aronov~\cite{AA79}. The correction to DOS
in the lowest order of perturbation theory was found to be more singular
than in the 2D case, $\delta \nu/\nu_0 \sim -1/\sqrt{\vert
  \epsilon\vert\tau}$.  Working beyond the perturbation approximation,
Nazarov~\cite{N} found that close to the Fermi surface DOS has an
exponential behavior, $\ln{(\nu/\nu_0)} \sim \vert \epsilon \vert^{-1}$ 
(this result was later obtained by Levitov and
Shytov~\cite{LS} in a different way).  Ballistic 1D conductors
were studied  by several authors~\cite{MG,Bo} and were shown to
have a power law behavior $\nu/\nu_0 \sim \vert \epsilon
\vert^\alpha$ of the tunneling DOS.  The crossover between diffusive
and ballistic regimes as well as the temperature behavior of DOS in
multi-wall carbon nanotubes were recently studied in Ref.~\onlinecite{MAG}
under the assumption that electrons reside on the outermost shell
only.

Here we study the zero-bias anomaly due to dynamically screened inter- and
intra-shell Coulomb interaction in two- and one-dimensional layered
systems and its effects on the tunneling DOS.  In Section~\ref{sec:2d} the
system of two-dimensional layers is analyzed, with both regimes of
diffusive and ballistic in-plane electron motion considered.  In
Section~\ref{sec:carbon} we discuss the zero-bias anomaly in multi-wall
nanotubes assuming that the doping electrons are distributed uniformly
across the shells.

\section{System of two-dimensional electron layers}
\label{sec:2d}

Let us consider a semi-infinite system of identical conducting
two-dimensional layers separated by the distance $d$ (see the
Figure in Appendix). The tunneling electron from TEM propagates within the
upper layer, and the inter-layer tunneling is neglected.  The
properties of an isotropic two-dimensional electron system are
described by the in-plane Fermi velocity $v$ and the
electron-impurity scattering rate $1/\tau$.  The presence of
internal layers is important as they contribute to the screening
of Coulomb interaction.
\par
The first order perturbation correction to the tunneling DOS (see Refs.\
\onlinecite{AA,RAG,KhR}) of the 2D conductor at zero temperature has the form
\begin{eqnarray}
\label{nu}
\frac{\delta \nu (\epsilon)}{\nu_0}=
\int\limits_{\vert \epsilon \vert}^{\epsilon_F}
d\omega {\cal V}(\omega),\\
\label{correction}
{\cal V}(\omega)= \Im
\int\limits_0^{\infty} \frac{qdq}{2\pi^2}
\frac{(\omega+i/\tau) U(\omega,q)\Gamma^2(\omega,q)}
{[(\omega+i/\tau)^2-q^2v^2]^{3/2}},
\end{eqnarray}
where $\nu_0$ is the thermodynamic two-dimensional density of states
$\nu_0=m/\pi$, counting both spin directions.  The electron-impurity
vertex function is given by
\begin{equation}
\label{vertex}
\Gamma^{-1}(\omega,q) = 1-\frac{i/\tau}{[(\omega+i/\tau)^2-q^2v^2]^{1/2}}.
\end{equation}
The function $U(\omega, q)$ stands for the Coulomb interaction of two
electrons residing in the outermost plane and screened by the infinite
number of conducting layers. To find this function, we consider the
Coulomb potential $\phi (\omega, {\bf r})$ created by the tunneling
electron located in the outermost plane $z=0$. It satisfies the Poisson
equation, that has the following form in the Fourier representation with
respect to the in-plane coordinates,
\begin{eqnarray}
\label{poisson}
\left( \frac{d^2}{dz^2} -q^2 +4\pi e^2 \Pi (q,\omega) \sum_{n=0}^{\infty}
\delta(z-nd) \right) \phi (\omega,q,z)\nonumber\\
= 4\pi e \delta(z),
\end{eqnarray}
where the last term in the brackets describes the polarization charge
induced in the system of 2D layers, and the polarization operator of a
single 2D electron layer is,
\begin{equation}
\label{equation}
\Pi (\omega,q) = \nu_0 ~ \frac{\omega+i/\tau-
[(\omega+i/\tau)^2-q^2v^2]^{1/2}}{[(\omega+i/\tau)^2-q^2v^2]^{1/2}-
i/\tau}.
\end{equation}
For the solution of Eq.~(\ref{poisson}) we refer the reader to the
Appendix.  We obtain,
\begin{equation}
\label{u}
U(\omega,q) \equiv -e\phi(\omega,q,0) = \frac{4\pi e^2 \sinh{qd}}{q
\left( e^{kd}-e^{-qd} \right)},
\end{equation}
where $k$ is given by the solution of the equation
\begin{equation}
\label{k}
\cosh{kd}=\cosh{qd}-K(\omega,q) \sinh{qd},
\end{equation}
having a non-negative real part $\mbox{Re}~k \ge 0$. Here
$K(\omega,q)={2\pi e^2 \Pi(\omega,q)}/{q}$ is the factor describing the
dynamical screening of an external charge by a single metallic layer.
When the wavelength decreases $qd \rightarrow \infty$, Eq.~(\ref{u})
gives
\begin{equation}
\label{uu}
U(\omega,q) = \frac{2\pi e^2}{q-2\pi e^2\Pi(\omega,q)},
\end{equation}
and the conventional expression for the screened interaction in a single
2D layer is recovered. For static interactions Eq.~(\ref{uu}) gives
$U(\omega \to 0,q) = 2\pi e^2/(q+\kappa)$, where $\kappa=2\pi e^2 \nu_0$
is the inverse static screening length. In what follows we assume that
$\kappa d \gg 1$. This condition ensures that different layers are (at
least for low frequencies) weakly coupled. In the opposite limit $\kappa d
\ll 1$ the system could be treated effectively as the 3d metal with the
cylindric Fermi surface. This regime is beyond the scope of our paper.

\subsection{Ballistic motion}
\label{bal}

First we consider the ballistic limit when the tunneling bias exceeds the
scattering rate $\epsilon \gg 1/\tau$.  The main contribution to the
integral in Eq.~(\ref{correction}) comes from two regions: 1)
high-frequency plasmon region, $qv \ll \omega$; 2) particle-hole continuum
$qv > \omega$.

1) Using the high frequency ($qv \ll \omega$) approximation for the
polarization operator Eq.~(\ref{equation}),
$\Pi(\omega,q)=\nu_0q^2v^2/2\Omega^2$, where we introduced the notation
$\Omega^2=\omega (\omega+i/\tau)$, and solving equation (\ref{k}) with
respect to $e^{kd}$, we get
\begin{eqnarray}
{\cal V}_1(\omega)&=&
\frac{1}{4\pi\epsilon_F\omega^2} \Im \int\limits_{0}^{q_\omega}
\frac{dq}{q}(e^{2qd}-1)\nonumber \\
&& \times
\left(\Omega^2-\omega_+ \omega_- -
\sqrt{(\Omega^2-\omega_+^2)(\Omega^2-\omega_-^2)}\right).
\label{low}
\end{eqnarray}
Here $\omega_\pm (q)$ represents the upper/lower boundaries of the
plasmon continuum (of an infinite system of layers $z=0,\pm 1, \pm 2,...$)
that can be parametrized by the wavenumber $q_\perp$ 
 in the $z-$direction,
\begin{equation}
\omega^2_{\pm}(q)=\kappa \frac{qv^2}{2} \left.
\begin{cases}{ \coth{\frac{qd}{2}}\cr
    \tanh{\frac{qd}{2}} }
\end{cases} \right\}
\end{equation}
The mode $\omega_+(q)$ corresponds to the uniform charge
distribution across all layers, $q_\perp =0$, while the other mode,
$\omega_-(q)$, describes the the alternating charge in adjacent
planes ($q_\perp=\pi/d$). The former mode has a frequency gap
$\epsilon^* =v (\kappa/d)^{1/2}$,  and at $qd \to 0$ gives the
usual three-dimensional plasmon (in anisotropic metal).  At $qd
\rightarrow \infty$, both branches tend to the usual plasmon
spectrum of a two-dimensional electron gas. This is illustrated by
Fig.~1.\\
  \narrowtext{
\begin{figure}
\begin{center}
  \epsfxsize=7cm \epsfbox{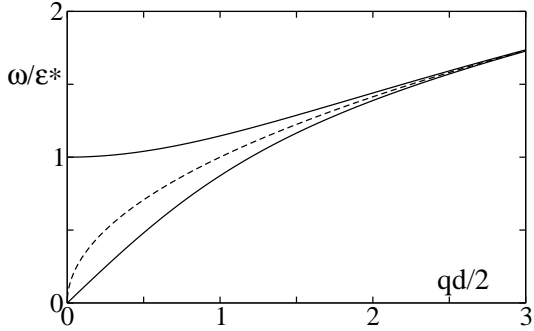}
\end{center}
\caption{The plasmon spectrum of an infinite number of 2D
metallic layers is shown. Dashed line shows the plasmon spectrum
of a single 2D layer. Plasmons of different layers interact with
each other thus creating a band (inner area between solid lines).
The upper solid line $\omega = \omega_+(q)$ represents the upper
boundary $\omega = \omega_+(q)$ of the plasmon continuum, with the
uniform charge distribution across the layers ($q_{\perp}=0$)
while the lower solid line marks the lower boundary corresponding
to the alternating charge in adjacent layers ($q_\perp=\pi/d$).}
\label{fig1}
\end{figure} }
We chose the momentum
cutoff $q_\omega$ in Eq.~(\ref{low}) to be larger than the characteristic
momenta of plasmon modes but still less than the momentum of particle-hole
excitations, i.e. $q_\omega \ll \omega/v$.

When the frequency is high $\omega > \epsilon^*$, the integral
over the momentum is dominated by large values of $q >1/d$ where
$\omega_+$ and $\omega_-$ converge exponentially,
$\omega_{\pm}^2=\kappa q v^2 [\frac{1}{2}\pm \exp{(-qd)}]$.
Simple calculations give,
\begin{equation}
\label{nu_1}
{\cal V}_1 (\omega) = -\frac{1}{2\epsilon_F}
-\frac{1}{2\pi \epsilon_F \omega \tau}
\ln{\frac{\kappa q_\omega v^2}{\omega^2}}.
\end{equation}

2) High transferred momenta $q \sim \omega/v \gg q_\omega$ present the
contribution of particle-hole pairs. Under the condition $\kappa d \gg 1$,
the second term in the right-hand side of Eq.~(\ref{k}) always exceeds
the first term (overscreened interaction) 
and the interaction takes the form,
\begin{equation}
\label{u_high}
U(\omega, q)= \frac{1}{\nu_0}
\frac{\sqrt{(\omega+i/\tau)^2-q^2v^2}-i/\tau}{\sqrt{(\omega+i/\tau)^2-
q^2v^2}-   \omega-  i/\tau}.
\end{equation}
Substituting Eqs.\ (\ref{u_high}) and (\ref{vertex}) into
Eq.~(\ref{correction}) and integrating from $q_\omega$ to infinity we find

\begin{equation}
\label{nu2}
{\cal V}_2 (\omega) = \frac{1}{4\epsilon_F}
-\frac{1}{2\pi \epsilon_F \omega \tau}
\ln{\frac{\omega}{q_\omega v}}.
\end{equation}
By adding this expression to the plasmon contribution (\ref{nu_1}) we obtain
the following expression for the total spectral weight function,
\begin{eqnarray}
{\cal V} (\omega)&=& {\cal V}_1 (\omega) +{\cal V}_2 (\omega) \nonumber \\
&=& -\frac{1}{4\epsilon_F}
-\frac{1}{2\pi \epsilon_F \omega \tau}
\ln{\frac{\kappa v}{\omega}}, ~~~ \omega > \epsilon^*.
\label{tot-high}
\end{eqnarray}

The results (\ref{nu_1}) and (\ref{tot-high}) hold as long as the
frequency exceeds the plasmon gap $\epsilon^*$. In this large frequency and
large wave number $qd>1$  region the plasmons have virtually no dispersion along the
$z$-direction due to weak interaction of electron densities on different
layers.  At lower frequencies and wave numbers, $\omega < \epsilon^*$ and
$qd<1$,  electron densities on different layers interact strongly, and the
plasmon spectrum acquires a significant dispersion along the
$z$-direction. In this region we may approximate $\omega^2_- = \kappa
d q^2 v^2/4$, $\omega^2_+=\kappa v^2/d$, $\exp{(2qd)} = 1+ 2qd$, and take
the inverse interlayer distance as the momentum cutoff $q_\omega \sim
2/d$. Taking the integral in Eq.~(\ref{correction}) explicitly and
extracting the imaginary part, we find instead of Eq.~(\ref{nu_1}),
\begin{equation}
\label{nu11}
{\cal V}_1 (\omega) = -\frac{1}{4\epsilon_F}
-\frac{1}{2\pi \epsilon_F \omega \tau}
\ln{\frac{\epsilon^*}{\omega}}.
\end{equation}
The high-momentum (particle-hole) contribution does not depend (as long as
$\kappa d \gg 1$) on the distance between layers. This can be readily seen
from Eq.~(\ref{u_high}) that does not contain $d$. This is quite natural
as the particle-hole pairs do not induce long-range oscillations of
electric field.  Therefore, we can take the old expression (\ref{nu2})
but with the new cutoff $q_\omega \sim 2/d$, and obtain
\begin{equation}
\label{tot-low}
{\cal V} (\omega) =-
\frac{1}{2\pi \epsilon_F \omega \tau}
\ln{\frac{\kappa v}{\epsilon^*}}, ~~ \omega < \epsilon^*.
\end{equation}
Integrating the expressions (\ref{tot-high}) and (\ref{tot-low}) 
with the frequency according
to Eq.~(\ref{nu}), we found the perturbation correction to the tunneling
DOS in the ballistic regime
\begin{eqnarray}
\frac{\delta \nu_{b} (\epsilon)}{\nu_0} &=&
\frac{\mbox{max} (\vert \epsilon\vert,\epsilon^*)-
\epsilon_F}{4\epsilon_F} \nonumber \\
&&-\frac{1}{4\pi \epsilon_F \tau}
\ln{[\frac{\kappa v}{\mbox{max}(\vert\epsilon\vert,\epsilon^*)}]} ~
\ln{[\frac{\kappa v}{\epsilon^2}\mbox{max}(\vert\epsilon\vert,\epsilon^*)]}.
\label{ball}
\end{eqnarray}
We observe that the correction (\ref{ball}) is less singular near the
Fermi surface than in the case of a single 2D layer.

Let us point out that in the limit $d\rightarrow \infty$ our
result differs from the formula of Khveshchenko and
Reizer,~\cite{KhR} as the leading contribution (the first term of
Eq.~(\ref{ball})) is twice as small as their result. The
difference is due to the contribution from particle-hole
excitations, ${\cal V}_2(\omega)$ in Eq.~(\ref{tot-high}).

\subsection{Diffusive motion}

Let us now address the regime of strong impurity scattering, $\epsilon \ll
1/\tau$, when the motion of the tunneling electron is diffusive and the
polarization operator has the form, $\Pi(\omega,q)=-\nu_0D
q^2/(Dq^2-i\omega)$, here $D=v^2\tau/2$ is the diffusion coefficient.

When $\omega > \tilde\epsilon = D\kappa/d$, the main contribution to the
integral (\ref{correction}) comes from the region of transferred momenta
$\omega/D\kappa \sim q< \sqrt{\omega/D}$, while for $\omega <
\tilde\epsilon$ the principal contribution is due to $\sqrt{\omega
  /D\kappa d} <q <\sqrt{\omega/D}$.  In this region the interaction takes
the form $$
U(\omega, q) = -\frac{i \omega}{\nu_0 D q^2},
$$
and we obtain
\begin{equation}
\label{diff}
{\cal V}(\omega) = -\frac{1}{4\pi^2 \nu_0 D\omega}
\ln{\frac{D\kappa^2}{\mbox{max}(\omega,\tilde\epsilon)}}.
\end{equation}
For a sufficiently dirty system $\tilde\epsilon \tau <1$. In this case
integrating over the frequency we obtain for the diffusive DOS correction
at $ \vert\epsilon \vert > \tilde\epsilon$,
\begin{equation}
\label{aal}
\frac{\delta \nu_d (\epsilon)}{\nu_0} =\frac{1}{8\pi^2\nu_0
  D}\ln{({D^2\kappa^4\tau}/{\vert\epsilon\vert})}
\ln{(\vert\epsilon\vert\tau)}+
\frac{\delta \nu_b (1/\tau)}{\nu_0},
\end{equation}
where the first term is the well-known
double logarithmic correction of Altshuler,
Aronov, and Lee~\cite{AAL} and the second term is the constant 
contribution
of ballistic frequencies $\omega > 1/\tau$ discussed in the Section
\ref{bal}.

For $\vert \epsilon \vert < \tilde\epsilon$, we have instead of Eq.~(\ref{aal})
\begin{equation}
\label{difdos}
\frac{\delta \nu_d (\epsilon)}{\nu_0} =\frac{1}{4\pi^2\nu_0 D}\left(
  \ln{(\kappa d)} \ln{(\vert\epsilon\vert
\tau)}+\frac{\ln^2{\tilde\epsilon \tau}}{2}
   \right)+
\frac{\delta \nu_b (1/\tau)}{\nu_0}.
\end{equation}
and the first logarithm in Eq.~(\ref{aal}) is cut off.
For a relatively clean system with $\tilde\epsilon \tau >1 $ (i.e. such that
$l^2 >d/\kappa$) the second term in the brackets in this equation is
absent.

Comparing Eq.~(\ref{difdos}) with the result by Zuzin~\cite{Z} for
the DOS correction for a 2D electron system screened by a bulk
metal, we can observe that in the diffusive regime at low energies
the role of an infinite set of 2D electron layers $z=d,2d,3d...$
is equivalent to a bulk metal screen located at a distance $d/2$
from the outermost plane.

\subsection{Interlayer tunneling}
\label{tunneling}

 Throughout the analysis performed above we
disregarded completely all possible tunneling transitions between
different layers. We now consider the corrections to the
density of states originating from interlayer tunneling and
establish conditions when such processes could be neglected.
We treat them perturbatively, in the lowest order in the tunneling
Hamiltonian,
\begin{equation}
\hat{H}_t=\frac{t}{2} \sum_{i} \int \frac{d^2p}{(2\pi)^2}
[\hat\psi^\dagger_{i+1}({\bf p}) + \hat\psi^\dagger_{i-1}({\bf
q})] \hat\psi_i ({\bf p}),
\end{equation}
which conserves the in-plane momentum during tunneling. Having in
mind mainly applications to the system with strong in-plane
scattering (like high-$T_c$ materials) we focus here on the
diffusive transport regime $\epsilon \ll  1/\tau$. The corrections
to the density of states to the lowest order in the inter-layer
tunneling amplitude $t$ are shown in Fig.\ 2. Of the six diagrams
drawn here the more important ones are the diagrams e) and f) that
contain four diffusons each. They correspond to the tunneling
corrections to the screened Coulomb interaction rather than to
corrections to Green's function of the tunneling electron
itself, which are given by the diagrams a) -- d).
  \narrowtext{
\begin{figure}
\begin{center}
  \epsfxsize=8cm \epsfbox{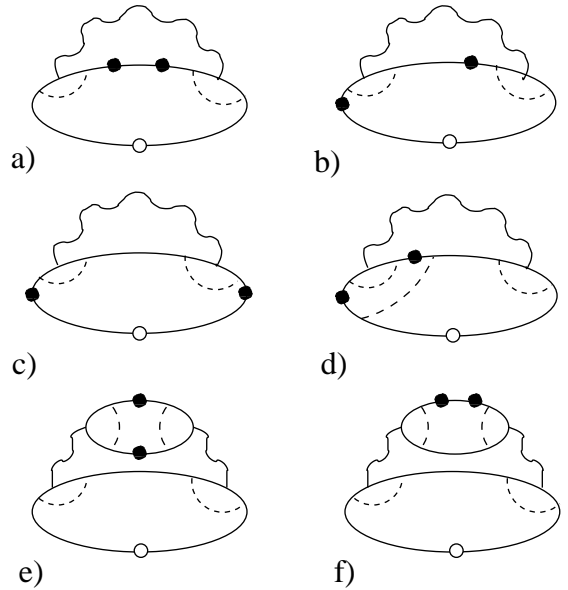}
\end{center}
\caption{ The diagrammatic representation of the correction to the
single-particle density of states from the inter-layer tunneling
processes to the lowest order of perturbation theory in the
tunneling amplitude $t$. The electron from the microscope probe
tunnels into white circles. Solid lines stand for the Green's
functions, wavy lines denote the dynamically screened Coulomb
interaction, black dots represent the tunneling matrix elements,
and the dashed lines are the impurity ladders (diffusons).}
\label{fig2 }
\end{figure} }
To compute these diagrams one has to know the interaction
$U_{nm}(\omega,q)$ between electrons residing on arbitrary layers
$n$ and $m$. As finding such a general expression seems to be
quite a cumbersome task for a semi-infinite system of layers, we
utilize the corresponding expressions for the infinite system to
obtain a qualitative estimate of the effect. It has the form (see
Appendix),
\begin{eqnarray}
\label{Unm} U_{nm}(\omega,q) &&= U_{00}(\omega, q) e^{-\vert n- m
\vert kd}, \nonumber\\
U_{00} (\omega,q) &=&\frac{2\pi e^2 \sinh{qd}}{q \sinh{kd}}.
\end{eqnarray}
The diagrams e) and f) are computed to yield for the tunneling
correction,
\begin{equation}
\label{tunn} {\cal V}_t (\omega) = t^2 \nu_0 \omega\tau ~ \Re
\int\frac{qdq}{2\pi^2} \frac{U^2_{00}(\omega,q) \tanh{kd/2}}
{(Dq^2-i\omega)^4}.
\end{equation}
If the frequency is large $\omega > \tilde\epsilon$, the main
contribution to the integral in Eq.\ (\ref{tunn}) arrives from the
interval  $1/d < q <\omega/\kappa D$, where we can 
approximate $ U_{00}
\tanh{kd/2} \simeq \kappa^2/\nu_0^2 q^2$, to obtain
\begin{equation}
\label{tunhigh} \frac{\delta \nu_t(\epsilon)}{\nu_0} =\frac{e^4
t^2\nu_0\tau}{\epsilon^2} \ln{\frac{\vert\epsilon\vert}{\tilde\epsilon}}.
\end{equation}
For smaller frequencies $\omega< \tilde\epsilon $, the leading
contribution comes from the integral $0<q<1/d$ and reads,
\begin{equation}
\label{tunlow} \frac{\delta \nu_t(\epsilon)}{\nu_0} =-\frac{e^4
t^2\nu_0\tau}{2\vert\epsilon\vert \tilde\epsilon}.
\end{equation}
It is worth noting that the inter-layer correction changes its
sign around the point $ \vert\epsilon \vert \sim \tilde\epsilon$ and in fact
lead to {\it increase} in the density of states for $ \vert\epsilon  \vert
>\tilde\epsilon$.
Comparing the expressions (\ref{tunhigh}) and (\ref{tunlow}) with
the above formulas (\ref{aal}) and (\ref{difdos}) we observe that
inter-layer tunneling correction is small provided that $t \ll
 \vert\epsilon \vert/(\epsilon_F\tau)$.
This much stronger condition than the  one that might be
naively expected ($t\ll  \vert\epsilon \vert$) is due to the enhancement
of  the electron-electron interaction by 
the inter-layer tunneling.
When this condition is violated the
summation of a  wider class of diagrams
 with all possible transitions
of diagrams is necessary.

\section{Multiwall nanotubes}
\label{sec:carbon}

In this section we consider the interaction suppression of tunneling DOS in
multiwall carbon nanotubes. A multiwall nanotube is built up of $M$
concentric carbon tubes (or shells) each  of which can be obtained by
rolling a graphite sheet into a cylinder.

In tunneling experiments~\cite{} the
tunneling current is believed to propagate along the outermost shell only
while the tunneling between the shells is largely suppressed. Thus the role of
electrons in the inner shells is reduced to merely screening the
interaction between the outer shell electrons.

The electron band structure of a single metallic carbon nanotube consists
of $N$ (this number is usually of the order of a few tens and depends on
the doping level and gate voltage) 
conducting sub-bands $\epsilon_n (k)$ characterized
by the Fermi velocities along the tube axis $v_n$ and around the
circumference $v_{\perp n}$. Each of the $M$
shells has, in general, its own
band structure $\epsilon_n (k)$. The electrons are scattered between
different bands (but mostly within the same tube) by impurities, lattice
imperfections and incommensurate potential of neighboring tubes.  Not all
the tubes are necessarily metallic, some of them could well be insulating.
Experimental evidence of the internal structure of MWNT is usually not
easy to obtain and leaves some room for speculation.  In our previous
paper~\cite{MAG} we studied the scenario of the dopants residing in the
outermost shell only, thus neglecting possible interaction with electrons
from internal shells.  In the present paper we study a different aspect of
the problem concentrating on the effects of the finite number $M$ of
conducting shells on the tunneling DOS of a MWNT.  To simplify the problem
we consider an approximation in which all $M$ shells have the same
band structure and the same doping level.

Unlike the case of the 2D conductors considered above, in the
one-dimensional wires the contribution of plasmon frequencies $\omega \gg
qv$, logarithmically exceeds  that of the electron-hole interval $\omega < qv$.
Therefore the weight function can be written in a
simple form (compare this
with Eqs.~(\ref{correction}) and (\ref{vertex})),
\begin{equation}
\label{correction2}
{\cal V}(\omega) = \frac{1}{\omega^2}
\Im \sum_{q_m} \int\limits_{-\infty}^{\infty} \frac{dq}{2\pi^2}
U_{00}(\omega,q,q_m),
\end{equation}
where the sum is taken over the quantized transverse momentum $q_m = m/R_0$,
$m=0,\pm 1, \pm 2...$ The function ${U}_{00}(\omega,q,q_m)$ should be
understood as the $00$ element of the matrix ${U}_{ij}(\omega,{\bf q})$
which represents the dynamically screened Coulomb interaction of an
electron in shell $i$ with an electron in shell $j$ and satisfies the
matrix equation
\begin{equation}
\label{Coulomb}
\hat{U}(\omega,{\bf q}) = \hat{V}({\bf q})+
\hat{V}({\bf q}) ~\hat{\Pi} (\omega,{\bf q})~ \hat{U}(\omega,{\bf q}),
\end{equation}
where $\Pi_{ij}(\omega,q,q_m)$ is the polarization operator which,
according to our assumptions, is proportional to the unit matrix
\begin{equation}
\label{pol}
\Pi_{ij}(\omega,q,q_m)= \delta_{ij} \Pi = \delta_{ij}  \nu_1
\frac{q^2 v^2_{\parallel}+q_m^2 v_\perp^2}
{\omega (\omega+i/\tau)},
\end{equation}
here we introduced the one-dimensional density of states $\nu_1 =\sum_n
(\pi v_n)^{-1}$ and the average squares of the longitudinal, $
v_{\parallel}^2$ and the transverse, $v_{\perp}^2$ electron velocities,
$$v_{\parallel}^2=\frac{\sum_n v_n }{\sum_n v_n^{-1}}, ~~~~~
v_{\perp}^2=\frac{ \sum_n v_n^{-1} v^2_{\perp n}} { \sum_n v_n^{-1}}.$$
In
Eq.~(\ref{Coulomb}) $V_{ij} (q,m)$ denotes the bare Coulomb potential,
\begin{equation}
\label{bare}
V_{ij} (q,q_m)=\frac{2e^2}{\pi}\int_0^{\pi} d\phi K_0 \left(qR_{ij}\right)
\cos{m\phi}.
\end{equation}
In this equation we introduced the notations
$R^2_{ik}({\phi})=R_i^2+R^2_k-2R_i R_k \cos{\phi}$,
with $R_i$ being the radii of concentric shells forming the MWNT ($i=0$ is
the external shell, $i=M-1$ the innermost shell).

\par To proceed further  with Eqs.~(\ref{Coulomb}-\ref{bare}) we assume that
the radius of the $j$-th shell is a linear function of its number $R_j=R_0
(1-j\xi)$.  In the long-range limit $qR_0 \ll 1$, we obtain from
Eq.~(\ref{bare}) for the bare interaction,
\begin{eqnarray}
  \label{vij}
\frac{V_{ij}(q,q_m)}{e^2}=\begin{cases}{
\frac{1}{\vert m \vert}
\left(\frac{1-\xi~\mbox{max} (i,j)}{1-\xi~ \mbox{min}(i,j)}\right)^{\vert
  m \vert}, & $m \neq 0$, \cr
 \ln{\frac{\beta}{(q R_0)^{2}}}
+2\xi~\mbox{min}(i,j), & $m=0$.}
\end{cases}
\end{eqnarray}
Here $\beta = 4e^{-2{\cal C}} \simeq 1.26$, with ${\cal C}$ being the
Euler constant.  While the second line in Eq.~(\ref{vij}) holds provided
$\xi \ll 1$, the first line utilizes only the approximation $qR_0 \ll 1$.

We use the set of eigenvectors $w_i^{(k)}(q)$ and eigenvalues $V_k(q,m)$
of the bare interaction matrix (\ref{bare}) to write the screened
Coulomb interaction in the form
\begin{equation}
  \label{eq:eigen}
  U_{ij}(\omega,q,q_m)=\sum_{k}\frac{w_i^{(k)}(m)w_j^{(k)}(m)}{V_k^{-1}(q,q_m)-
    \Pi(\omega,q,q_m)}.
\end{equation}
The eigenvalues $V_k(q,q_m)$ determine the spectrum of collective plasmon
excitations $\omega^{(k)}_m (q)$ through the poles of the interaction
$U_{ij}(\omega,q,q_m)$ in Eq.~(\ref{eq:eigen}), and the eigenvectors
$w_i^{(k)}(m)$ determine the distribution of charge between different
shells in these plasmon oscillations.  In the limit $qR_0 \gg 1$ one
can approximate sum with the integral
and recover the usual two-dimensional Coulomb potential (\ref{uu}).

Typically, MWNTs exhibit ballistic transport around the circumference $l
\equiv v\tau \sim 10-100 ~nm > R$.  At energies $\epsilon$ below the
Thouless energy $v/R$ for the transport around the circumference the
contribution of the $m\neq 0$ terms is non-singular due to the gap in
their spectrum and is only weakly dependent on $\epsilon$.  The $m=0$
plasmons, on the other hand, are gapless. Their contribution to
Eq.~(\ref{correction2}) continues to depend on $\epsilon$ and has a
singularity at $\epsilon \to 0$.  Therefore, concentrating on the low
energy $\epsilon < v/R$ dependence of DOS we can neglect the non-singular
contribution of the $m\neq 0$ plasmons and retain only the $m=0$ term in
Eq.~(\ref{correction2}).

The index $k$ labels plasmon modes and equals the number of nodes
in the charge distribution in the $k-$th plasmon across the
section of a multi-shell nanotube.  The mode $k=0$ is
characterized by the uniform distribution of the oscillating
charge and corresponds to the logarithmic eigenvalue, $V_0 (q)
\simeq e^2 M\ln {\frac{\beta}{(q R_0)^{2}}}$, with $w_i^{(0)}
\simeq 1/\sqrt{M}$.
All other $M-1$ modes correspond to $q$-independent eigenvalues of
the bare interaction, $V_k = 2\xi e^2\alpha_k$ and therefore have
sound-like spectrum.  The coefficients $\alpha_k$ are to be
computed numerically. They range roughly from a few tenths to a
few units, e.g. for $M=10$ we obtain $\alpha_k = 10.0 ; 2.6; 1.2;
0.72;0.50;0.38;0.31;0.28; 0.26$. The components $w_0^{(k)}$ of the
eigenvectors are also computed numerically. As the eigenvalues of
the matrix (\ref{eq:eigen}) depend (at most) only logarithmically
on the momentum $q$, we can perform the integral in
Eq.~(\ref{correction2}) taking these eigenvalues at characteristic
plasmon momenta, $q^{2} \sim \omega (\omega+i/\tau)/N v^2$,
\begin{equation}
\label{nu1} {\cal V}(\omega) = -\sqrt{\frac{g}{N}} f_M(\omega) ~
\Re\frac{\sqrt{\omega+i/\tau}}{\omega^{3/2}},
\end{equation}
where we defined the average Fermi velocity $\bar{v}=\sum_n v_n/N$
and the dimensionless coupling strength $g=e^2/(2\pi \bar{v})$ in
a single channel. The function $f_M (\epsilon)$ is given by the
$00$-element of the matrix $\hat{V}^{1/2}$,
\begin{eqnarray}
\label{coeff}
f_M(\omega) &=&\sum_k (w_0^{(k)})^2 \sqrt{V_k/2e^2} \nonumber\\
&&= \frac{1}{\sqrt{M}} \ln^{1/2}{\frac{\sqrt{NM} \bar{v}}{R
\sqrt{\vert\omega(\omega+i/\tau^{-1})\vert}}} +\gamma_M
\sqrt{\xi},
\end{eqnarray}
the last term representing the contribution of the $M-1$ sound-like
plasmons, with values of $\gamma_M$ given in the table.

\begin{center}
\begin{tabular}{| c | c | c | c | c | c |}
\hline
$M$ & 1 & 2 & 3 & 5 & 10 \\
\hline
~$\gamma_M$~ & ~~0~~ &~ 0.35~ & ~0.60~ & ~0.84~ & ~1.38~~\\
\hline
\end{tabular}
\end{center}

The contribution of the $k=0$ plasmon decreases with the number of
tubes $M$ due to the screening by internal shells, while the
contribution of sound-like plasmons increases roughly linearly
with $M$. Since the number of shells may not exceed $1/\xi$ the
second term in Eq.~(\ref{coeff}) is never greater than the first
one, and at most becomes comparable to it at $M\approx 1/\xi$. The
approximation $\omega > qv$ used above assumes that the case of
strong interaction is realized, $\xi \alpha_k > 1/4 e^2\nu_0 \sim
1/4N$. In the opposite limit $\xi \alpha_k < 1/4N$, the
contribution of the $k-$th
plasmon mode is weak and may be treated perturbatively.

Substituting the expression (\ref{nu1}) into Eq.\
(\ref{correction2}) and evaluating the frequency integral we obtain
the following perturbation correction to the tunneling DOS of an
$M$-shell multi-wall nanotube,
\begin{eqnarray}
\label{dosper} \frac{\delta \nu_M (\epsilon)}{\nu_1} =
-\sqrt{\frac{g}{N}}\nonumber\\
\times \begin{cases}{ \left(
\frac{\sqrt{2}}{\sqrt{\vert \epsilon \vert
\tau}}f_M( \vert\epsilon \vert)+ f_M(\tau^{-1})
\ln{\frac{\sqrt{NM} \bar{v}\tau}{R}} \right), &  $ \vert\epsilon  
\vert <
1/\tau$, \cr f_M(\epsilon) \ln{\frac{\sqrt{NM} \bar{v}}{ \vert
\epsilon  \vert
R}}, & $\vert \epsilon \vert> 1/\tau$ . }\end{cases}
\end{eqnarray}

We now discuss how the results obtained for multi-wall nanotubes
could be extended to the very vicinity of the Fermi surface where
the perturbation correction (\ref{dosper}) diverges and the
nonperturbative approach is required. This question was addressed
for a single-wall metallic nanotube in Ref.\ \onlinecite{MAG} that
utilized the phase approximation first proposed by Nazarov\cite{N}
and further developed in Ref.~\cite{KA}.
Here we concentrate on the dependence of the DOS on the number of
shells $M$ skipping the derivation that could be found in Ref.\
\onlinecite{MAG}. The result for the ballistic regime,
$ \vert\epsilon \vert > 1/\tau$ mainly utilizes the plain exponentiated
perturbative result (\ref{dosper}),
\begin{equation}
\label{balexp} \nu(\epsilon) \propto \exp \left\{
-\sqrt{\frac{g}{N}} \ln{\frac{\bar{v}}{R \vert\epsilon \vert}}
\left(\frac{1}{\sqrt{M}} \ln^{1/2} \frac{\bar{v}}{R \vert\epsilon \vert}
+\gamma_M \sqrt{\xi} \right) \right\}.
\end{equation}
The diffusive regime of lower energies, $ \vert\epsilon \vert < 1/\tau$, 
is more
tricky. We obtain,
\begin{equation}
\label{difexp} \nu(\epsilon) \propto \exp \left\{
-\frac{\epsilon_0}{ \vert\epsilon \vert} \left(\frac{1}{\sqrt{M}} \ln^{1/2}
\frac{\bar{v}}{R \vert\epsilon \vert} +\gamma_M \sqrt{\xi} \right)^2 \right\},
\end{equation}
where $\epsilon_0 = \pi g/2N\tau$. This result holds provided that
the argument of an exponent is large and the temperature small
enough $T< \epsilon^2/\epsilon_0$. For higher temperatures
$\epsilon^2/\epsilon_0 < T < \epsilon$ the tunneling DOS acquires
the form,
\begin{equation}
\label{diftemp} \nu(T) \propto \exp \left\{ -1.07
\sqrt{\frac{\epsilon_0}{T}} \left(\frac{1}{\sqrt{M}}
\ln^{1/2} \frac{\bar{v}}{R\sqrt{T \epsilon_0}} +\gamma_M
\sqrt{\xi} \right)^2 \right\}.
\end{equation}
The crossover between the low-temperature (\ref{difexp}) and the
high-temperature (\ref{diftemp}) regimes can be shown
to depend on the energy through the
dimensionless variable $ \vert \epsilon  \vert/\sqrt{\epsilon_0 T}$ only.
We omit the derivation of the corresponding scaling function
which is quite straightforward
and follows the same route as in the case
when  logarithms in the exponents are screened off
by the presence of a metal shield, see Ref.~\onlinecite{MAG}. 


\section{discussion}
\label{sec:discussion}

We considered the zero-bias anomaly in in the tunneling density of
states in layered 2D and quasi-1D materials with dynamically screened
Coulomb interaction. The theory presented above is applicable 
to high-T$_c$ materials and semiconductor heterostructures
as well as to the multi-wall carbon nanotubes. We showed that the
presence of many conducting shells in 2D systems weakens the 
singularity in the DOS correction at the Fermi level: for energies
below $\epsilon^*=\kappa D/d$ in diffusive systems or below the
plasmon gap $v(\kappa/d)^{1/2}$ in ballistic systems 
the DOS correction becomes 
logarithmic $\propto \ln  \vert\epsilon \vert$, rather than double-logarithmic 
$\propto\ln^2 \vert\epsilon \vert$ as in the case of a single layer.

In addition to the logarithmic terms mentioned above that
originate from the broad region of the transferred frequency and
momentum there are also terms coming from the regions around the
plasmon and particle-hole singularities. These contribute
correction to DOS $\mbox{max}(\vert \epsilon
\vert,\epsilon^*)/4\epsilon_F$ that is not singular but dominant in
the wide range of energies $\vert \epsilon \vert > \ln
(\epsilon_F\tau)/\tau$.
The contribution of the particle-hole continuum is especially
interesting in 2D ballistic systems. It turns out that the 
contribution of the particle-hole continuum to the DOS correction, 
which corresponds to the region $\omega \leq qv$ in the interaction 
frequency/momentum transfer, of the same order as the contribution
originating from the plasmon pole but has a {\it different} sign.
Namely, the particle-hole continuum {\it increases} the density of
states.  The total effect is still a 
suppression of DOS for low energies as the plasmon contribution is
always larger than that of the particle-hole continuum (twice as large
for a single 2D layer). 

In derivation of DOS in 2D systems we assumed that
the ground state is the free electron model. 
It is possible, however, 
to incorporate also the Fermi liquid interaction in the calculations
of the tunneling DOS.
In particular, the 
Fermi liquid theory predicts that thermodynamic density
of states is renormalized
$\nu_0 \to Z \nu_0$ by 
a $Z$-factor ($0<Z<1$) describing a step in the quasiparticle
partition function.
We obtain however that the relative correction $\delta \nu/\nu_0
=\vert \epsilon \vert /4\epsilon_F$
due to Coulomb interaction is in fact {\it universal} as 
it is  independent of the
Fermi liquid interaction strength $Z$.

The validity condition for  neglecting the inter-shell
tunneling was found for the diffusive regime. It was shown that
instead of the naively expected requirement of the tunneling time
$t^{-1}$ being greater than the characteristic measurement time
$\epsilon^{-1}$, the much more stringent condition 
$t^{-1} \gg \epsilon_F \tau/\epsilon \gg \epsilon^{-1}$ 
must be satisfied.

With respect to multi-wall carbon nanotubes our analysis was
concentrated on the dependence of DOS on the number $M$ of coaxial
shells for both the regimes of perturbative and strong
suppression. We found that this dependence can be described by a
single function $\gamma_M$ found numerically. The energy and
temperature dependence of DOS found in Ref.~\onlinecite{MAG} for a
nanotube with a single conducting shell is preserved for an arbitrary
number of shells.

\section{acknowledgement}
\label{sec:acknowledgement} The authors are grateful to
I.~Beloborodov, A.~Chubukov, L.~Glazman, A.~Larkin, M.~Pustilnik 
and M.~Turlakov for valuable
discussions. We thank K.~Efetov for warm hospitality at Bochum
University where part of this work was performed. This research
was sponsored by the Grants DMR-9984002 and BSF-9800338 and by the
A.P.~Sloan and the Packard Foundations. E.\ M.\ also thanks the
Russian Foundation for Basic Research, Grant 01-02-16211.

\appendix
\section{Solution of the Poisson equation}

To solve the Poisson equation (\ref{equation}) we first derive the
Green function for the infinite problem,
\begin{equation}
\label{green}
\left( \frac{d^2}{dz^2} - q^2 +4\pi e^2 \Pi \sum_{n=-\infty}^{\infty}
\delta(z-nd) \right) G (z,z')= \delta(z-z'),
\end{equation}
with the boundary condition $G(z,z') \to 0$ when $\vert z-z' \vert \to
\infty.$ The homogeneous solutions of Eq.~(\ref{green}) are easily
written in terms of quasiperiodic Bloch functions,
\begin{eqnarray}
\label{homog}
\psi_{\pm}(z) = e^{\pm knd}\left(\sinh{[q(z-(n+1)d]}-e^{\pm kd}\right.\nonumber\\
\left. \times  \sinh{[q(z-nd)]} \right), ~~~
nd < z <(n+1)d
\end{eqnarray}
with $k$ given by the solution of the equation
\begin{equation}
\label{k1}
\cosh{kd}=\cosh{qd}-\frac{2\pi e^2 \Pi}{q} \sinh{qd}
\end{equation}
having a positive real part, $\mbox{Re}~k \ge 0$.  This choice means that
the function $\psi_-(z)$ decreases and the function $\psi_+(z)$ increases
with $z$ increasing.  The Green function can then be written in the
conventional form
\begin{equation}
\label{green1}
G(z,z')= -\frac{1}{2q\sinh{qd}\sinh{kd}} \begin{cases}{
\psi_-(z) \psi_+(z'), & $z>z'$, \cr
\psi_+(z) \psi_-(z'), & $z<z'$. } \end{cases}
\end{equation}
To find the solution for the semi-infinite problem, Eq.~(\ref{poisson}),
we impose a fictitious charge $Q$ located at $z=-d/2$ (see Fig.\ 3).
  \narrowtext{
\begin{figure}
\begin{center}
  \epsfxsize=8cm \epsfbox{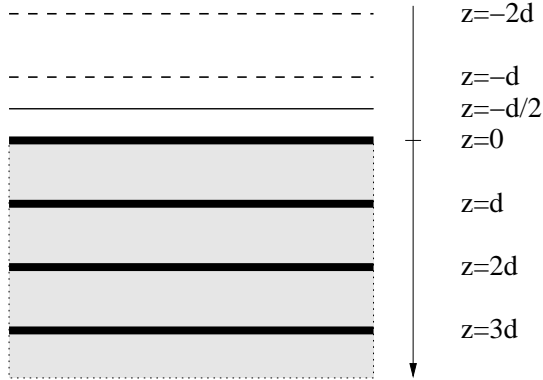}
\end{center}
\caption{ The crystal consisting of a semi-infinite system of 2D
metallic layers ($z=0,d,2d,..$) is shown. The thin solid line at
$z=-d/2$ hosts a fictitious (image) charge $Q$ that accounts for
the absent $z=-d,-2d,..$ layers.} \label{fig3 }
\end{figure} }
 This charge has to
be found from the condition that the total potential of the
electronic charge $-e$ and the fictitious charge $Q$
\begin{equation}
\label{phi} \phi(z)=4\pi [e ~ G(z,0)-Q ~G(z,-d/2)]
\end{equation}
contains only the exponent $e^{qz}$ for the negative values of the
coordinate, $-d/2<z<0$.

This condition ensures that the electric field in
the outside region ($z<0$) decays exponentially with the distance from the
outermost plane, with the fictitious charge, therefore, taking care of the
absent ($z=-dn$, $n=1,2,...$) planes.  We obtain the following value of
the fictitious charge
\begin{equation}
\label{charge} Q= e
\frac{\psi_-(0)}{\psi_+(-d/2)}\frac{e^{kd}-e^{qd}}{e^{-kd}-e^{qd}}.
\end{equation}
Substituting Eq.~(\ref{charge}) into Eq.~(\ref{phi}) and making use of
Eqs.\ (\ref{homog}) and (\ref{green1}) we obtain the final expression
(\ref{u}).

In Section \ref{tunneling} the formula for interaction $U_{nm}$ of two
electrons residing at different layers $z=nd$ and $z=md$ in an
infinite system is used. To obtain it one can write $U_{nm} =
-4\pi e^2 G(nd,md)$, which leads to the expression (\ref{Unm}).

\end{multicols}
\end{document}